**Nanyang Technological University**

**School of Communication and Information**

# ADOPTION OF THE E-GOVERNMENT IN THE REPUBLIC OF KAZAKHSTAN

## Report on Critical Inquiry

By

Yerlan Amanbek, Ilyas Balgayev, Kanat Batyrkhanov, Margaret Tan

May 2009

# ADOPTION OF E-GOVERNMENT IN THE REPUBLIC OF KAZAKHSTAN

**Yerlan Amanbek, Ilyas Balgayev, Kanat Batyrkhanov, Margaret Tan**
Nanyang Technological University
Singapore 637718
Email : {w070057; yerl0001; baty0001}@ntu.edu.sg

**ABSTRACT**
This paper identifies factors that influence Kazakhstan's e-Government portal usage. It determines challenges encountered by citizens while using the portal. Targeted respondents for the web-based questionnaire survey were the citizens of Kazakhstan. The technology acceptance model was used as a methodology to measure attitude towards portal usage. In addition, this paper discusses the barriers which were experienced by the respondents and can prevent the successful adoption of e-Government initiative. The results of the analysis demonstrate that awareness among citizens is high, i.e. the majority of people have visited it at least once and they perceive the portal to be useful, but only a limited percentage of citizens' use it on the regular basis. Further, paper can be used to help IT managers of the portal to improve management of informational content and maintain a more effective adoption among citizens.

## INTRODUCTION

The governments and public organizations over the world have taken advantages of the Internet and information and communication technologies (ICT). It can be seen from the fact that governments of industrialized and developing countries utilized and made investments into e-services and e-Government (Choudrie, Ghinea, & Weerakkody, 2004).

There are many definitions of the e-Government concept. According to the European Information Society "e-Government is the use of information and communication technology in public administrations combined with organizational change and new skills in order to improve public services, democratic processes and strengthen support to the public policies" ("European Information Society," 2009). Carter and Belanger (2005) explained e-Government as an information technology that enables and improves the efficiency with which government services are provided to citizens, employees, businesses and agencies. With the help of e-Government, state organizations get many benefits in terms of time and cost reduction as well as general growth of effectiveness of civil agencies while offering services to citizens.

Many industrialized countries, such as the USA, Singapore, Australia, and Netherlands have succeeded in development and implementation of the e-Government concept. Singapore developed a consistent and integrated way to computerize the government sector. In 1980 it realized the significance of IT in establishing its economy (Tan, 1998), and today Singapore's e-Government program stands among the most developed in the world. Clear vision and strong leadership have become the main factors, which underlie the success of e-Government in Singapore. The government of Australia understood the importance of one-stop portal e-Government either. It devised an integrated approach to deliver e-services in what is widely known as a "single window" (Reffat, 2006). The Single Window approach is used, when all the required documents completed together, and supposed to be shared among different departments with high level of transparency, with difficulty for someone to hide unjust behavior (Apostolov, 2008).

While industrialized countries introduced the use of e-Government, developing countries also started to undertake e-Government initiatives adapting existing previously created models within their own context. Many countries are trying to introduce e-Government concepts with high enthusiasm, but they do not always consider prerequisites and concurrent management issues (Hakikur, 2007).

According to Heeks (2004), 35 percent of e-Government initiatives failed, 50 percent partially failed and only 15 percent were successful. However, several developing countries, such as Brazil, India and Chile have prospered in their implementation of e-Government initiatives (Ndou, 2004). These examples show that governments of developing nations can effectively use and appropriately employ the advantages of ICT. It is recognized that there is significant gap between advanced and developing countries in e-Government development but it is believed that developing countries can learn from industrialized countries (Song, 2006).

**E-Government of the Republic of Kazakhstan**

The Republic of Kazakhstan, which is a developing country, is also trying to build an informational society based on using ICT. According to the Nation Address of President of the Republic of Kazakhstan (*Toward a Competitive Kazakhstan, a Competitive Economy, and a Competitive Nation*, 2004), the Agency of Information and Communication has developed National Program for Developing e-Government in the Republic of Kazakhstan for Year 2005-2007. This is one of the serious steps in building the highly technologically developed government.

The main objective of introducing e-Government is to increase the quality and effectiveness of public administration and public services. E-Government aims to automate the state agencies' activities. This could allow citizens to use highly demanded e-services.

Currently Kazakhstan's government is struggling with challenges such as a bureaucratic pattern in governance, complication of redundancies in public sectors, shortage in coordination and proper information dissemination among governmental sectors. Most of these problems are planned to be eliminated with the help of effective use of ICT through establishing one centralized database where all relevant information regarding a particular citizen will be electronically accessible to authorized parties whenever it is needed.

The creation of e-Government in Kazakhstan is a long-term process, which is planned to be implemented through three sequential stages:

The initial, informative (from 2005 to 2006) stage included single assess point to all information resources of the state bodies. During this stage, initiators conducted orientation towards citizens, provided information regarding organizations' needs, proceeding from the life and business events. Within the framework of the stage, information about services of all state bodies had been placed in the portal to the end of 2006. Generally, it provided citizens with government related information.

The second, interactive (from 2007) stage has been implementing mechanisms of identification and authorization of the users, citizens reception services, and mobile version of the portal. After completion of this stage, citizens will be able to send applications and inquiries in order to get specific information or complete wide range of procedures through the Internet.

The third, transactional (from 2008) stage, fulfils mechanisms of citizens' access to payment of public services, through the portal. It will make possible financial transactions by means of the integration of payment gateway and banking system. After completion of this stage, people will be able to make payments for the public services through the Internet.

In order to make the realization of e-Government to be feasible, only developing and introducing of web-based applications is insufficient. There is also need in so called "informational society", i.e. people who possess sufficient computer literacy and inexpensive affordable access to ICT, who are able to interact with the government through the Internet regardless of their geographical location. In order to do this, the Agency of Information and Communication has developed a program on "Reducing informational inequality program for years 2007-2009". This program, which totally costs around US$1 million, aims to increase the number of active Internet users by 20 percent of Kazakhstan's population by the end of 2009. The main objectives of the program are to reduce the cost of Internet connections, provide social computers at an affordable prices, form interest in ICT among Kazakhstani citizens (AIC, 2006). The Agency distributed numerous booklets among population, organized "round tables" about e-Government, offered free training for citizens.

Nevertheless, awareness among citizens of this initiative is low. As it was stated in the article ("Colossal amount of money was spent on e-government realization," 2008), most of money was spent on technical aspects of the project and the Agency's efforts were not enough first hand. There are many reasons of unpopularity of the e-Government portal. One of the issues may be a level of trust of the population with respect to government electronic resource. It is striking that people tried to search for Kazakhstan e-Government portal and could not even register there. Even some city dwellers are not fully aware of what is going on in the portal. However, according to initiators of the project, even small villages are planned to be provided with special terminals with access to e-Government.

**Problem statement**

According UN every country has the e-Government readiness index ("Government readiness," 2008). The e-Government readiness index is a composite index comprising the web measure index, the telecommunication infrastructure index and the human capital index. In 2005, the survey showed that Kazakhstan was ranked in the 65th place with e-Government readiness index 0.4813, but in 2008, its rank dropped down to 81 and e-Government readiness index became 0.4743.

However, the government has allocated a colossal amount of money (about US$120 million) for the implementation of e-Government program (*National Program for Developing e-Government in the Republic of Kazakhstan for Year 2008-2010*, 2007), and US$100 million have already spent in 2008 ("Colossal amount of money was spent on e-government realization," 2008). A lot of money was spent, despite of this fact a little progress has been made in implementation of e-Government.

**Scope**

Since Kazakhstan e-Government portal is considered as a main gateway to all state departments, units and their electronic services we will focus on and investigate challenges in its usage. E-Government is completed partially and exists only in the first informative stage; therefore, we focused only on informative functionalities and features of the portal.

**Objective**

This paper investigates readiness and citizens' motivation to the portal. The main goal is to identify challenges and influencing factors encountered by citizens in using e-Government through the Internet technology.

## LITERATURE REVIEW

To understand the factors encountered by citizens, we chose the Technology Acceptance Model (TAM), which is used as a model to measure attitude toward technology adoption in multiple domains. Subsequent researches confirmed usefulness of TAM and its various extensions and revisions as a tool for investigating and predicting user information technology acceptance (Al-khateeb, 2007; Lin & Sheng, 2002; Liu, Chen, & Zhou, 2006; M. G. Morris & Dillon, 1997; van der Heijden, 2003).

AlAwadhi and Morris (2008) showed that the factors, which determine the adoption of e-Government services, identified in particular developed countries, could be transferable in the context of developing countries.

Original TAM, developed by Davis, describes a person's acceptance of a technology by his (or her) voluntary intentions toward using the technology (Davis, Bagozzi, & Warshaw, 1989). TAM predicts user acceptance based on the influence of two factors: perceived usefulness (PU) and perceived ease of use (PEOU). Figure 1 shows the original TAM, which states that user's perception of usefulness and ease of use determines attitude toward using (AT) the system, and further behavioral intention to use (BI) is determined by this attitude. According to the TAM, behavioral intention determines actual system use (AU). TAM proposes a direct relationship between perceived usefulness and behavioral intention to use.

TAM offers a relatively simple and cost-effective way to predict whether a system is actually used (Money & Turner, 2004).

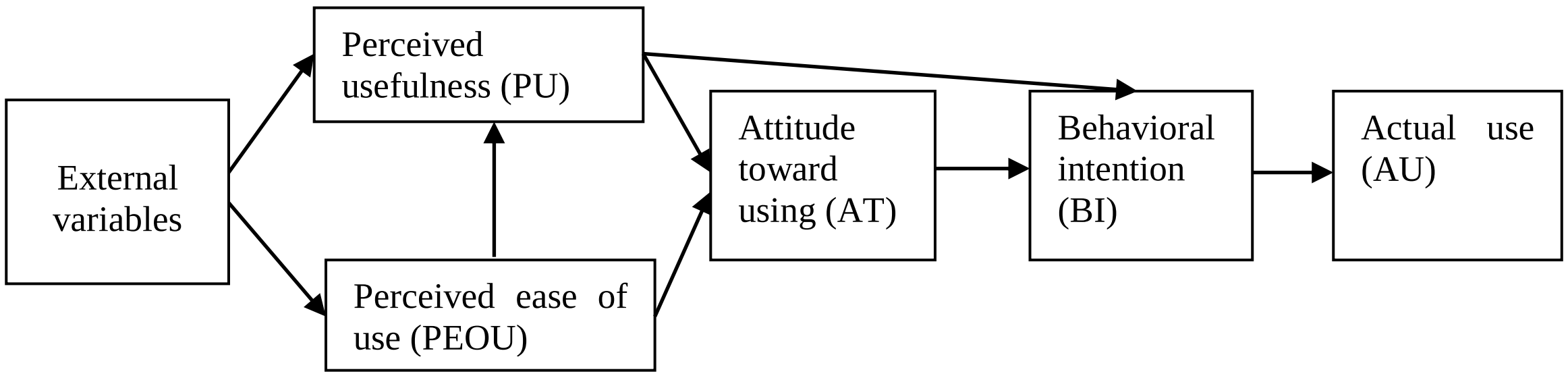

**Figure 1. Original TAM.**

Perceived usefulness is defined as the "subjective probability that using a specific application system will increase his or her job performance within an organizational context". Perceived ease of use stands for "the degree to which the user expects the target system to be free of efforts" (Davis et al., 1989). Attitude toward using is described as a personal feeling about performing the target behavior (Fishbein & Ajzen, 1975). Attitudes towards using is an object that influences intentions which, in turn, influence behavior with respect to the object (Ajzen & Fishbein, 1980). Behavioral intention could be explained as strengths on one's intentions to use the technology in the future.

For the context of technological acceptance, it is essential to measure attitude towards using and perceptions regarding the use of technology rather than attitude and perceptions directed towards the technology itself, because people could have a positive view about a technology without positive feeling towards its use (Yousafzai, Foxall, & Pallister, 2007).

Davis et al., (1989) proposed to revise the original TAM, which they claim as a more powerful model in prediction and explanation of user behavior. It based only on such constructs as PU, PEOU and BI. AT construct was removed, because of the partial mediation of the impact of perceptions on intention by attitude. Later research on the TAM tried to indicate that attitude may play a central mediating role for determining mandatory usage, but its direct relationship to behavioral intention was not supported (Adams, Nelson, & Todd, 1992; Jackson, Chow, & Robert, 1997). Performance is a key, intention forms on performance considerations rather than person's feelings, with respect to performing behavior (Taylor & Todd, 1995).

The revised TAM, shown in the Figure 2, has a strong relationship between perceptions and intention. It has been shown that attitude may exist to correlate itself with usage behavior (Yousafzai et al., 2007). Freedom of choice has a significant impact on person's acceptance of the innovation (Ard-Barton, 1988). For example, if some employees do not want to accept an innovation, they may even sabotage implementation. Brown, Massey, Montoya-weiss, and Burkman (2002) explained this reaction as a personal attitude, which drives employees in their form of hostility towards the technology.

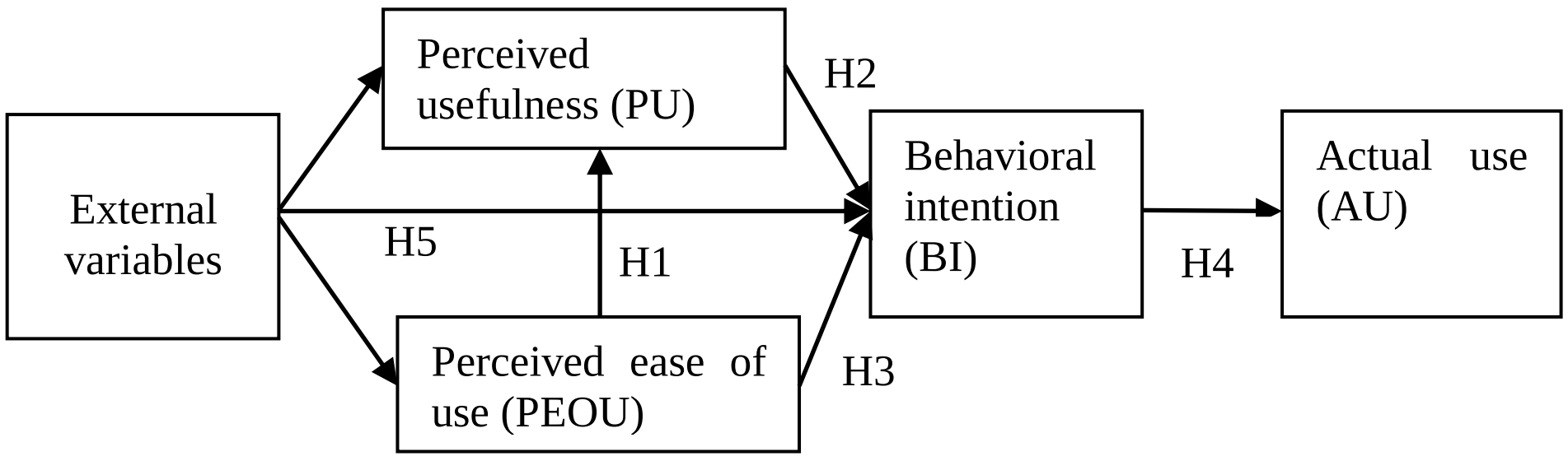

**Figure 2. Revised TAM.**

**Hypotheses**

According to Davis, perceived ease of use has a direct influence on perceived usefulness. Morris and Dillon (1997) mentioned that between two systems with similar features and functionalities, a user should find the one that is easier to use to be more useful. Davis states that if a user works more productively because of enhancement of ease of use, then he or she should work more productively overall.

*H1: Perceived ease of use has a positive influence on perceived usefulness*

TAM states that both PU and PEOU constructs have a significant impact on user's acceptance. Since the beginning, many studies have described PU to be more important than PEOU and that PEOU has no direct relationship to AU (Venkatesh, 1999). Davis found that in the early stages, PEOU has a stronger impact on intentions, but with time and experience, it was found that the effect is indirect, operating through PU (Davis et al., 1989). However, it would be fair to say that no amount of PEOU would compensate for low usefulness (Keil, Beranek, & Konsynski, 1995). The role of PEOU in TAM, has an equal or even stronger effect than role of PU on technology adoption (Igbaria, Zinatelli, Cragg, & Cavaye, 1997). For example, people intend to use the technology if they perceive it usefulness, but if the technology would not be easy to use, it could lead to avoidance of usage.

*H2: Perceived usefulness has a positive influence on behavioral intentions to use*

*H3: Perceived ease of use has a positive influence on behavioral intentions to use*

There is a direct and positive effect between behavioral intention to use the portal and actual use of the portal. Research on TAM has consistently demonstrated a strong empirical support for this relationship: intentions can accurately explain and predict actual behavior.

*H4: A behavioral intention to use has a positive influence on actual use*

Many studies have tested the effects of external variables on PEOU and PU (Hong, Thong, Wong, & Tam, 2002; Igbaria et al., 1997). Compeau, Higgins, and Huff (1999) and Morris and Venkatesh (2000) discussed independency of PEOU and PU from the effects of external variables. However, Agarwal and Prasad (1999) found that PEOU and PU totally mediated the impact of external variables on user' attitude and intention to use it. Theory of Reasoned Action (TRA) Fishbein and Ajzen (1975) states that external variables influence attitude only through user perceptions about the technology.

In our study, we used an academic level and occupation as the external variables that have an influence on behavioral intention to the use of the e-Government portal.

*H5: External factors (academic level and occupation) have a significant impact on behavioral intentions to use*

## METHODOLOGY

We conducted quantitative research through the web-based survey to get the perception towards the portal.

Pitkow and Recker (1995) noted the advantages of web based surveys such as structured responses, possibility to point-and-click responses, transfer of electronic medium for data and its comparison, visual representation of the questions, flexible time constraints and constructing questions in advance manner to reduce the number and complexity of questions presented to users.

We used *"SurveyMonkey.com"* web site to create and publish the questionnaire. This web site offers transformation of URLs into direct links to the web site, as it was suggested by van Selm and Jankowski (2006), would lead to minimization of additional actions on behalf of the respondents.

### Questionnaire design

The questionnaire consists of three sections and has 23 questions. The first section captures questions concerning the external factors that are associated with e-Government portal usage. The second section examines factors on perception of the e-Government portal through the prism of TAM. These factors consist of perceived ease of use, perceived usefulness, behavioral intention and actual use. The third section set out to capture general profile of respondents. Please see Appendix A for detailed information.

A seven point Likert scale was used, ranging from "1" - strongly agree to "7" - strongly disagree.

The survey lasted for 1 week from 12 to 19 March 2009. The questionnaire originally constructed in English, it was necessary to translate it into Kazakh and Russian, the local spoken languages. The questionnaire was designed to be short, unambiguous and easy for the respondents to complete. Constructs and statements relevant to the study were adopted from the a variety of sources mainly, past researches (Morris & Dillon, 1997; van der Heijden, 2003) and research papers with similar interests (Al-khateeb, 2007; Choudrie & Dwivedi, 2005; Lee & Lei, 2007; Lin & Sheng, 2002; Soury, Ziaee, & Shenassa, 2008) and modified to suit the research context.

**Sample**

The target respondents for the survey were Kazakhstan citizens. Invitations for participation in the survey were distributed among our ex-colleagues, friends, and friends of friends, family members.

**Data analysis**

The statistical package for the social services (SPSS) was used to analyze data and to determine an existence of relationships. We used the statistical techniques known as correlation and regression. The correlation coefficient measures strength and direction of the relationship. The closer the value of the correlation coefficient is to +1 or -1, the stronger the linear relationship is between two variables (Bluman, 2006).

Since the number of items for PU, PEOU and BI constructs was more than one, shown in the Table 1 and the Appendix B, Table B1, the principle components factor analysis was performed. Details are in the Appendix C, Tables C2, C3 and C4. Principle components factor analysis is a method for re-expressing multivariate data. It allows reorienting the data so that the first few dimensions account for as much of available information as possible. If there is substantial redundancy present in the data set, then it may be possible to account for most of the information in the original data set with a relatively small number of dimensions (Tabachnick & Fidell, 2000). In addition, we used a Kaiser's Varimax rotation, to achieve simple structure by focusing on the columns of the factor loadings matrix.

| Construct | Number of items | Question number |
|---|---|---|
| PU | 23 | Q9, 11 and 12 |
| PEOU | 24 | Q10, 13 and 14 |
| BI | 6 | Q15-16 |
| AU | 1 | Q8 |

**Table 1. Number of questions allocated for TAM constructs.**

Since the nature of key variables in hypotheses was different, two approaches – Linear regression and Independent Sample T-test were conducted to test existence of correlation. Details of variables for hypotheses are presented in the Table C1 of the Appendix C.

As the first approach, hypothesized relationships were tested by using regression analysis to maintain consistency with earlier studies. For our research, a stepwise multiple regression analysis technique is recommended to examine the influence of each predictor variables to the regression model (Hair, Anderson, Tatham, & Black, 1995). As one of the multiple regression types, linear regression was applied to facilitate the analysis. Particularly, linear regression analysis is a tool with several important applications. It is a way of testing hypotheses concerning the relationship between numerical variables and estimating the specific nature of such relationships (Carver & Nash, 2006). Linear regression investigates the relationship of predictor variables (independent variable) to outcome variables (dependent variable). The main difference between independent variable and dependent variable is that independent variable can be controlled or manipulated (Bluman, 2006).

Another approach was the T-test, which identifies the significance of difference between two means. One of the most commonly used T-tests is the Independent Samples T-test. We can use this test when we need to compare the mean of two independent samples on a given variable. Urdan (2005) suggested conducting the Independent Samples T-test, when one variable is categorical and another, dependent one is numeric.

## FINDINGS

### Respondents' overview

We sent out 300 emails and the response rate was 73 percent. 60 percent of respondents were males and 40 percent were females. Two age groups were represented by almost all respondents, the majority of respondents were from the first age category, 16 – 25 (71 percent), and second category was people of 26 – 35 years old (16 percent). This could be mostly explained by the method of the survey distribution which was non-random. In terms of education, 53 percent of respondents held a bachelor degree and whereas a master's degree was held by 50 percent of respondents. The number of PhDs who participated in the survey was low (2 percent). Results of demographic profile are presented in the Table 2.

| **Category** | | **Percentage** |
|---|---|---|
| **Gender** | Male | 59.6 |
| | Female | 40.4 |
| **Age** | 16-25 | 71.3 |
| | 26-35 | 16.4 |
| | 36-55 | 7.6 |
| | 56 and above | 4.7 |
| **Academic level** | High school | 15.2 |
| | University bachelor | 52.6 |
| | University master | 29.8 |
| | University PhD | 2.3 |

**Table 2. Demographic profile.**

People from a finance (19 percent) and IT (15 percent) background were involved actively but the percentage of students was the highest (32 percent). A salary range of 75001–150000 tenge (US$501–1000) had led the list with 31 percent; the second place was given for salary range of 150001–300000 tenge (US$1001–2000) with portion of 23 percent. Almost two third of respondents claimed Almaty (50 percent) and Astana (22 percent) as cities where they stay.

The result showed that the majority of respondents (51 percent) used the Internet more than ten hours per week and 22 percent used it from one to four hours accordingly. 52 percent of respondents accessed the internet from home, and 38 percent did from work. However, we had cases where both access points were stated (6 percent). Broadband connection type was the common answer (65 percent), while a dial-up connection was the second common answer with 30 percents. Most usual price for the internet connection was in range 0–4500 tenge (US$0–30) for 74 percent.

**Respondent's awareness of e-government portal**

It was found that 81 percent of participants were aware of the e-Government portal. From this set of people 69 percent visited the portal at least once. From those who answered for question "How many times do you visit the e-Government portal per week?", only 20 percent indicated that they used the portal on a regular basis, one or more times per week (Table 3 and Table 4). The top three informational sources that promoted the portal were web sites, TV and information from friends and family members. Figure 3 presents sources of how people knew about the portal.

| Question | Yes | No |
|---|---|---|
| **Have you ever heard about e-government portal ?** | 80.82% | 19.18% |
| **Have you ever visited the Kazakhstan's e-Government portal?** | 68.97% | 31.03% |

**Table 3. Awareness of the portal**

| Question | more than 1 times per week | less than 1 times per week |
|---|---|---|
| **How many times do you visit the e-Government portal per week?** | 20% | 80% |

**Table 4. Actual use.**

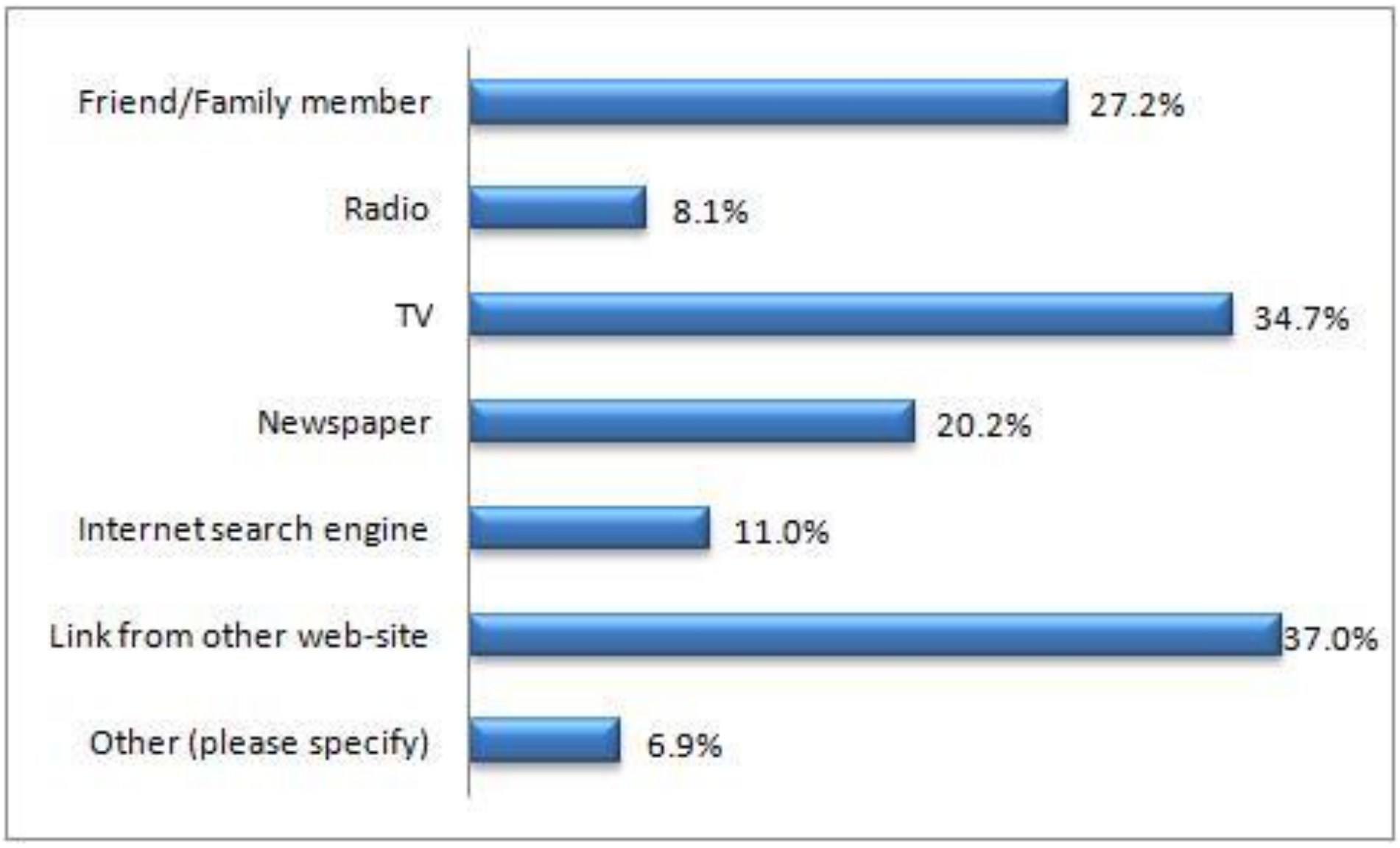

**Figure 3. Summary of the answers for question: "How did you know about Kazakhstan e-Government portal?"**

Since our respondents were young people who used Internet frequently, the web-sites became the most popular source (37 percent). In addition, the government pays attention to common mass media channels such as TV, radio and newspaper which were ranked highly (35 percent for TV). Viral marketing effect took place in promotion of e-Government portal when a person disseminate information among friends, friend's friends and family members (27 percent).

**Actual users**

An actual user group is respondents who used the e-Government portal regularly (more than 1 time per week) comprising 20 percent. They are indicated in table 4.

There was no significant difference between genders. Males held 53 percent and females comprised 47 percent. Both males and females were almost equally involved in the portal use. Most frequent people who used the portal were in range of 16-25 years old (47 percent), followed by age group of 26-35 with 29 percents. A number of respondents that used portal reduced with age increasing. In the context of Kazakhstan, elder people do not use the Internet and even computers as frequently as youth do. Citizens with a master degree comprised 42 percent and those who have bachelor degree set up 35 percent of the respondents. None of the PhDs used e-Government portal on a regular basis. The majority of actual users are academic degrees. It can be explained that content is interesting and useful for educated people but not much dedicated for others. People with bachelor and master degrees more widely use the internet and there is a higher level of awareness of e-Government among these groups of people. Historically Kazakhstani education system was derived from the USSR system, where scholars at the age of around 40-50 usually earned PhD degree. Assuming that majority of Kazakhstani PhD degree holders are elder than the age of 50, they may not be using the Internet expertly and prefer traditional methods of searching for information and interacting with government. Table 5 presents demographical profile of actual users.

| **Category** | | **Percentage** |
|---|---|---|
| **Gender** | Male | 53 |
| | Female | 47 |
| **Age** | 16-25 | 47 |
| | 26-35 | 29 |
| | 36-55 | 23 |
| **Academic level** | High school | 23 |
| | University bachelor | 35 |
| | University master | 41 |

**Table 5. Demographic profile of actual users.**

The majority of respondents who used the portal on a regular basis had broadband connection (65 percent). Prices for the access have become relatively lower comparing to the year 2007, when an unlimited dial-up plan was US$111 and 1.5 Mbps connection was US$3.355 (Anderson, 2007 ). It is clear that the availability of the internet became more affordable. Because, today Kazakhtelecom state owned telco offers broadband connection at the monthly price 1930 tenge (US$ 12.8).

The top three categories of occupation were government workers (35 percent), students (29 percent) and IT specialists (18 percent). None of the surveyed unemployed and sales people used the e-Governmental portal. Government workers might use the portal because it was directly related to their work. While using the portal, IT workers might find it useful and generally could be interested in e-Government initiative. Students were the major group of respondents of the survey and they could be interested in the education sector.

It was found that citizens of Astana used the e-Government portal more often than citizens of Almaty and other regions did. Astana is the capital city of the Republic of Kazakhstan and many Astana's citizens work in the government sector. Naturally, citizens of Astana are much closer to the government and more involved in governmental programs and initiatives.

Actual users were people mostly with income level of 75001-150000 tenge (US$501–1000) with 47 percent followed by the second group with income level of 15001-75000 tenge (US$100–500) comprising 35 percent. Together these two groups contained 82 percent and dominated over other groups. These groups determine the middle class of Kazakhstan citizens.

**Barriers in not using e-Government portal**

For those respondents, who had not used the portal (31%), we asked them for reasons why they were not using the portal. Please refer to Figure 4.

The majority of respondents (72 percent) stated that they did not think that they need to use the e-Government portal. The initiative might be new for the population and citizens have not got accustomed well to this concept. It could be explained from the point that it is much easier for people to get information through phone calls, mail request or face-to-face communication. Therefore, the government should encourage people to use it, through well-managed and well-promoted plans for breaking the line of resistance.

Another important aspect was absence of useful information (about 17 percent). It means that provided information was too generic and might not satisfy user needs. The portal provides external web links for details of particular matter, but they also could be insufficient, because participants said that they were not useful enough.

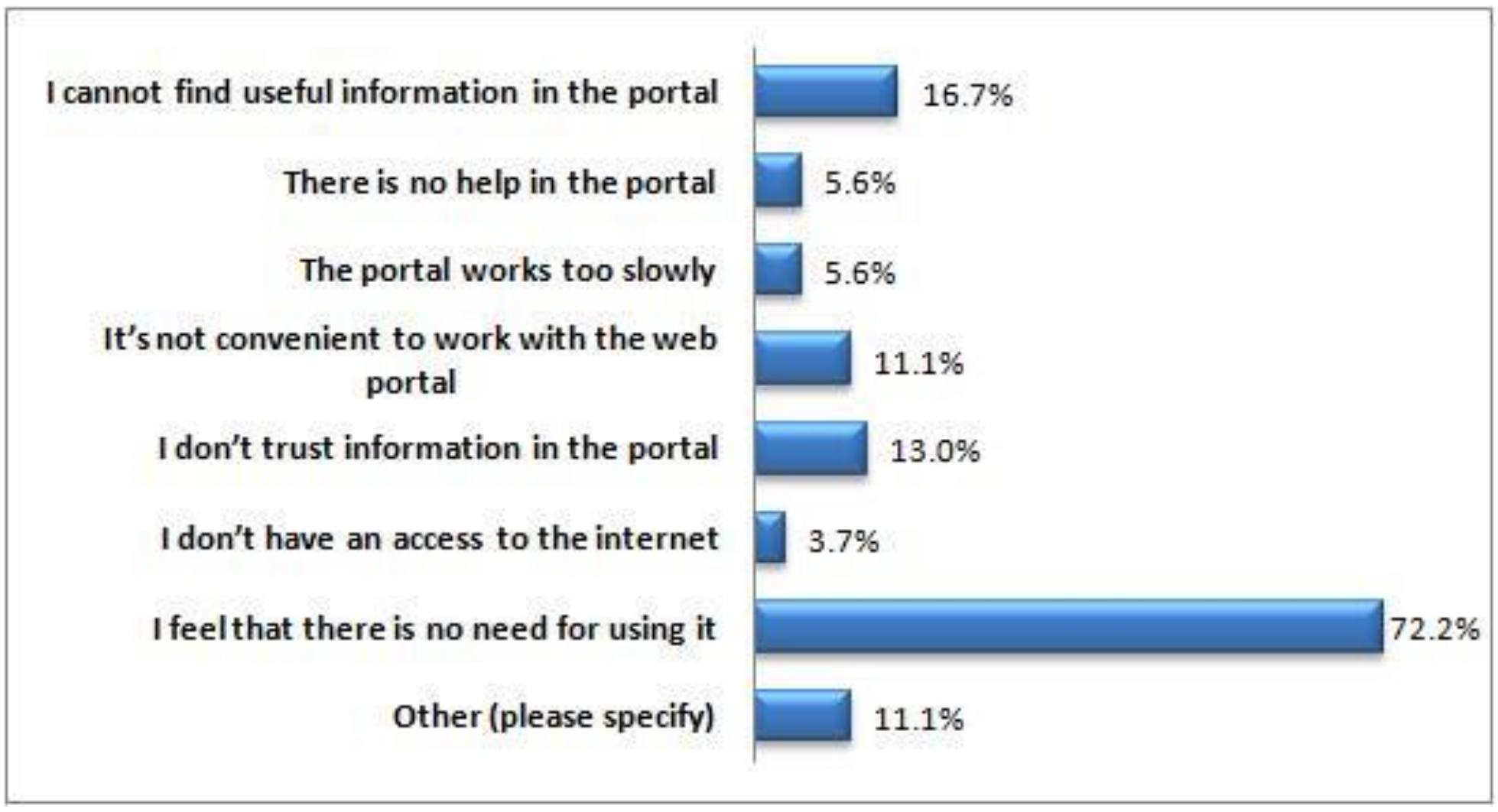

**Figure 4. Barriers for not using e-Government portal.**

It was interesting to note that 13 percent did not trust the information in the portal. It could be particularly explained by the fact that citizens are required to provide personal information (ID) in order to register as legal user. Recently, number of websites of different governmental departments was attacked by hackers. Therefore people might feel that their identity could be stolen and used in criminal actions. ("Kazakhstan: Five political parties report about the information terrorists to the public prosecution office," 2009)

**Content of e-Government portal**

We also investigated the most useful information sectors and information about services, their relevance and potential. Table 6 and 7 present a list of the top three informational sectors and information about services respectively. Full ranking tables are available in the Appendix E, in Tables E1, E2, E3 and E4.

In terms of sectors, education, health and employment were seen to be the three top most useful information sectors. The majority of respondents were students who felt that education and employment information sectors are important for them. However, health is the common aspect for any group of people.

| Sectors | PU | |
|---|---|---|
| | Mean | Std. Deviation |
| Education | 2.6 | 1.321 |
| Employment | 2.68 | 1.371 |
| Health | 2.73 | 1.394 |

**Table 6. Ranking of information sectors.**

| Services | PU | |
|---|---|---|
| | Mean | Std. Deviation |
| Airports | 2.98 | 1.537 |
| Municipal reference services | 2.98 | 1.705 |
| Railroad stations | 3.03 | 1.553 |

**Table 7. Ranking for information about services**.

The top three results based on usefulness for information about services were airports, municipal reference services and railroad stations. All these information about services implied transportation aspect which was interesting for citizens due to Kazakhstan large territory.

In addition, participants felt that portal was not convenient to work with. Many of functionalities were still under implementation, the content and structure of the portal could be changed over short period. It may confuse people and distract from further usage.

In fact, mostly respondents had a positive perception regarding the usefulness of the e-Government portal (PU mean value was 3.07 out of 7). However, the majority felt that the information did not fully satisfy their needs. Generally, respondents claimed that the portal was easy to use and they did not feel that they needed help while using it (PEOU mean value was 3.47 out of 7).

It was interesting to note that an overall intention of participants concerning e-Government portal usage was positive (BI mean value was 3.19 out of 7). Particularly participants indicated that they would use the portal to reduce time spent and get latest information and news. We face a contradiction in situation where majority intends to use e-Government portal, but in reality, they do not use it.

**Hypotheses**

Hypothesis 1 examined the link between perceived usefulness and perceived ease of use of e-Government portal. The respondents' value of usability was directly related to the value of usefulness of the portal. This hypothesis was accepted ($p<0.01$).

Even though the web portal provides useful information, if users would not be able to obtain it because of poor ease of use, eventually it might fail to satisfy their needs and users will not visit portal again. This empirical finding indicates that ease of use is the significant determinant for the portal usefulness.

Hypothesis 2 examined the link between perceived usefulness and behavioral intention; generally, participants intend to use e-Government portal in case if it satisfies their needs. Results of analysis demonstrated that participants had a positive perception on usefulness and their intention to use the portal was positive as well. This hypothesis was accepted ($p<0.01$).

Hypothesis 3 examined whether there is a positive influence on behavioral intention if user perceive portal to be easy to use. Hypothesis was rejected ($p= 0.08$, is greater than the acceptable level of 0.05). Although an overall intention was optimistic (mean value was 3.19 out of 7), perceived ease of use towards behavioral intention was not significantly correlated. Although, the e-Governmental portal is easy to use from users' point of view, it does not provide enough useful information. It appeared that troubles in finding needed information would lead to a resistance in the intention to use.

Hypothesis 4 examined the link between people's behavioral intention and actual use. The hypothesis was rejected ($p < 0.139$). People have a positive intention to use e-Government portal, but actual use is still remains low.

Hypothesis 5 examined whether there is an existence of significant correlation between behavior intention and external factors (occupation and academic level). BI variable was dependent on PU, PEOU, occupation, academic level variables. As a result, the Academic level variable and PEOU variable were excluded. The hypothesis was rejected. However, significant (5%) correlation between behavior intention and occupation, PU was found ($p<0.01$ for PU and $p<0.05$ for Occupation).

Tables D1 and D2 of the Appendix D provide detailed results of hypotheses calculations. The hypotheses tests were summarized in Table 8.

| | **Hypotheses** | **Result** |
|---|---|---|
| H1 | Perceived ease of use has a positive influence on perceived usefulness of e-Government portal | Accepted |
| H2 | Perceived usefulness has a positive influence on behavioral intentions to use e-Government portal | Accepted |
| H3 | Perceived ease of use has a positive influence on behavioral intentions to use e-Government portal | Rejected |
| H4 | A behavioral intention to use has a positive influence on actual use of e-Government portal | Rejected |
| H5 | External factors (Academic level and Occupation) have a significant impact on behavioral intentions to use e-Government portal | Rejected |

**Table 8. Summary of hypotheses testing.**

## CONCLUSION

The main purpose of the paper is to identify factors that influence the adoption of Kazakhstan e-Government portal using on the technology acceptance model.

Findings of this research present that people know about the e-Government portal, perceive it useful and easy to use, they intend to use it however actual use is low. Barriers affecting the adoption of this portal were identified. The most significant one was that people do not think that they need to use it. We suggest that the government should start promoting the importance of the portal through mass-media channels, especially TV, newspapers and publicizing information about it in reliable web based information sources. The most useful information sectors and information about sectors were identified.

People expect that the portal will solve the problem of bureaucracy and minimize time to process inquiries, which particularly belong to the interactive and transaction stages. However information provided in the portal does not match user's needs. Particularly respondents indicated that they cannot find information in the portal, and do not trust information in the portal which in turn affect intention to use.

The research confirmed that percived usefullness was an important factor in adoption of e-government portal. In order to increase adoption rate, developers of e-government portal should concern ways to provide usefulness in the portal.

Trust is another important factor for effective adoption of e-government. It should be considered for all stages in the further implementation of the portal. Therefore, the matter of trust should be established from the beginning, i.e. possessing up-to-date news and information from complete and trustable sources should be considered by managers of the portal.

Analysis shows that the e-Government portal could be treated as unsuccessful. However, it exists only on its first stage where problems are common. Giving an appropriate and timely response to them will minimize their occurrence in future. We hope that information and data provided in this paper will help in the implementation and maintenance of the e-Government portal.

## LIMITATIONS

The survey duration was only one week which is a short period. The survey employed a non-random convenience sample with a constraint of 219 participants. Collecting a significantly larger sample using an alternate survey modality and random sampling methods would be an expensive endeavor.

## APPENDIX A

### Questionnaire in English

We are a group of university students from Nanyang Technological University (Singapore) doing research on user's perception towards the adoption of Kazakhstan's e-Government portal (www.e.gov.kz). The information we collect will be used purely for academic purposes. All individual responses will be kept in strict confidence. Your participation is voluntary and you can withdraw at any time. This is an anonymous survey, which means that your name and contact particulars are collected only for the purpose of obtaining your consent to participate and will not be attached to your survey response.

We appreciate your time and effort in helping us to complete this survey. Please answer the questions openly and we seek your honest opinion.

Thank you,

Ilyas Balgayev, Yerlan Amanbek, Kanat Batyrkhanov

{w070057, yerl0001, baty0001} @ntu.edu.sg

| № | Question |
|---|---|
| 1 | How frequently do you use the internet per week? (Please choose one option)<br>☐ Less than 1 hour<br>☐ 1 – 4 hours<br>☐ 5 – 9 hours<br>☐ More than 10 hours |
| 2 | Where do you usually access the internet? (Please choose one option)<br>☐ Work<br>☐ Home<br>☐ Internet café<br>☐ Other, please indicate_________________________ |
| 3 | What is your internet connection type? (Please choose one option)<br>☐ Dial-up<br>☐ Broadband<br>☐ Satellite |
| 4 | How much do you pay for internet connection per month? (Please choose one option)<br>☐ 0 – 4500 tenge (US$30)<br>☐ 4501 tenge (US$30) – 7500 tenge (US$50)<br>☐ 7501 tenge (US$51) and above |
| 5 | Have you ever heard about the Kazakhstan e-Government portal?<br>☐ Yes<br>☐ No (Please go to the question 18) |
| 6 | How did you know about Kazakhstan e-Government portal? (Please select as many as necessary)<br>☐ Link from other web-site<br>☐ Internet search engine<br>☐ Newspaper<br>☐ TV<br>☐ Radio<br>☐ Friend/Family member<br>☐ Other, please indicate_________________________ |
| 7 | Have you ever visited the Kazakhstan's e-Government portal?<br>☐ Yes<br>☐ No (Please go to the question 17) |
| 8 | How many times do you visit the e-Government portal per week? (Please choose one option)<br>☐ Less than once<br>☐ 1 – 5 times<br>☐ 6 – 9 times<br>☐ 10 times and above |

The following questions require You to answer them in likert-scale format. All the items employed a seven-point scale for measurement, with anchors of "1" indicating "Strongly agree" and "7" indicating "Strongly disagree".

| № | Question | Strongly agree | Moderately agree | Slightly agree | Neutral | Slightly disagree | Moderately disagree | Strongly disagree |
|---|---|---|---|---|---|---|---|---|
| 9 | With reference to the Kazakhstan e-Government portal I find the content: | | | | | | | |
| | Current | | | | | | | |
| | Complete | | | | | | | |
| | Accurate | | | | | | | |
| | Non redundant | | | | | | | |
| | Useful | | | | | | | |
| 10 | With reference to the Kazakhstan e-Government portal I find … | | | | | | | |
| | it easy to understand | | | | | | | |
| | It easy to search for information | | | | | | | |
| | It frustrating | | | | | | | |
| | I need help while using it | | | | | | | |
| | There are too many links that confuse me | | | | | | | |
| | It easy to use | | | | | | | |
| 11 | I find the information in following sectors useful: | | | | | | | |
| | Education | | | | | | | |
| | Health | | | | | | | |
| | Employment | | | | | | | |
| | Culture | | | | | | | |
| | Sport | | | | | | | |
| | Tourism | | | | | | | |
| 12 | I find the Information on following government and private services useful: | | | | | | | |
| | Airlines | | | | | | | |
| | Motor transport | | | | | | | |
| | Airports | | | | | | | |
| | Railroad stations | | | | | | | |
| | Hotels | | | | | | | |
| | Transportation | | | | | | | |
| | Post | | | | | | | |
| | Internet | | | | | | | |
| | Telecommunications | | | | | | | |
| | Employment | | | | | | | |
| | Municipal reference services | | | | | | | |
| | Legal services | | | | | | | |
| 13 | I can find relevant and appopriate information that I need easily in the following sectors: | | | | | | | |
| | Education | | | | | | | |
| | Health | | | | | | | |
| | Employment | | | | | | | |
| | Culture | | | | | | | |
| | Sport | | | | | | | |
| | Tourism | | | | | | | |
| 14 | I find using following information on government and private services easy to use: | | | | | | | |
| | Airlines | | | | | | | |
| | Motor transport | | | | | | | |
| | Airports | | | | | | | |
| | Railroad stations | | | | | | | |
| | Hotels | | | | | | | |
| | Transportation | | | | | | | |
| | Post | | | | | | | |
| | Internet | | | | | | | |
| | Telecommunications | | | | | | | |
| | Employment | | | | | | | |
| | Municipal reference services | | | | | | | |
| | Legal services | | | | | | | |
| 15 | I would like to use e-Government web portal because it helps me: | | | | | | | |
| | To reduce my expenses | | | | | | | |
| | To reduce my time spent | | | | | | | |
| | To minimize bureaucracy | | | | | | | |
| | To get latest information and news | | | | | | | |
| | To get quick responses for inquiries | | | | | | | |
| 16 | I would not use the portal as I am currently using other sources for the information I need (e.g. informal web sites, official resources and mass media channels) | | | | | | | |
| 17 | What are the reasons for not using/visiting Kazakhstan e-Government web portal? (Please select as many as necessary)<br>☐ I cannot find useful information in the portal<br>☐ There is no help in the portal<br>☐ The portal works too slowly<br>☐ It’s not convenient to work with the web portal<br>☐ I don’t trust information in the portal<br>☐ I don’t have an access to the internet<br>☐ I feel that there is no need for using it<br>☐ Other, please indicate ____________________ | | | | | | | |

| | | |
|---|---|---|
| **Demographic profile** | | |
| 18 | Gender | ☐ Male ☐ Female |
| 19 | Agee | ☐ 16 – 25 ☐ 26 – 35<br>☐ 36 – 55 ☐ 56 and above |
| 20 | Academic level | ☐ High school ☐ University Bachelor<br>☐ University Master ☐ University PhD |
| 21 | Place of staying | ☐ Almaty ☐ Astana<br>☐ Other, please indicate_____________ |
| 22 | Occupation | ☐ Government ☐ Finance<br>☐ Engineering ☐ IT<br>☐ Sale ☐ Student<br>☐ Unemployed ☐<br>☐ Other, please indicate_____________ |
| 23 | Salary range | ☐ under 15000 tenge (US$100)<br>☐ 15001-75000 tenge (US$100-500)<br>☐ 75001-150000 tenge(US$501-1000)<br>☐ 150001-300000tenge (US$1001-US$2000)<br>☐ Above 300000 tenge (US$2000) |

**Questionnaire in Kazakh**

Біз, Наньянг Технологиялық Университетінің студенттері, «Қолданушылардың Қазақстан Республикасының «Электрондық үкімет» (www.e.gov.kz) порталын кабылдауы» атты зерттеу жұмысын жасаудамыз. Осы зерттеу жұмысындағы сауалдамадан алынған мәлеметтер тек қана академиялық мақсатта қолданылады. Барлық Сіз берген жауаптар қатаң конфеденциалды түрде сақталынады. Осы сауалдамада Сіз еркін турде қатысасыз, және кез-келген уақытта сурактардан бас тарта аласыз. Бұл сауалдама анонимдік болып табылады, яғни сіздің атыңыз және байланыс контакттыңыз (сіз берген жағдайда) сауалнамада қатысу келисіміңізді ғана білідіреді.

Біз сізді болген уақытыңыз үшін жоғары бағалаймыз, сұрақтарға ашық және шын жауап беруіңізді өтінеміз.

Ізгі ықыласпен,

Балгаев Ильяс, Аманбек Ерлан, Батырханов Канат

{w070057, yerl0001, baty0001} @ntu.edu.sg

| № | **Сұрақ** |
|---|---|
| 1 | Сіз аптасына интернетті қанша уақыт колданасыз? (Бір жауапты таңдаңыз.)<br>☐ 1 сағаттан кем<br>☐ 1-4 сағат<br>☐ 5-9 сағат<br>☐ 10 сағат және жоғары |
| 2 | Сіз Интернетті әдетте қай жерде қолданасыз? (Бір жауапты таңдаңыз.)<br>☐ Жұмыста<br>☐ Үйде<br>☐ Интернет кафеде<br>☐ Басқасы, нұсқаңыз ________________________ |
| 3 | Сіз қандай интернет байланыс түрін қолданасыз? (Бір жауапты таңдаңыз.)<br>☐ Коммутацияланған телефон желілері бойынша қатынау (Dial-Up)<br>☐ Жылдам желі<br>☐ Спутниқалық |
| 4 | Интернет байланысы үшін айына қанша төлейсіз? (Бір жауапты таңдаңыз.)<br>☐ 0 – 4500 теңге (US$30)<br>☐ 4501 теңге (US$30) – 7500 теңге (US$50)<br>☐ 7501 теңге (US$51) және жоғары |
| 5 | Қазақстан «Электрондық үкімет» порталы туралы естідіңіз бе?<br>☐ Иә<br>☐ Жоқ (18-ші сұраққа барыңыз) |

| | | | | | | | | |
|---|---|---|---|---|---|---|---|---|
| 6 | Қазақстан «Электрондық Үкімет» порталы туралы қай жерден білдіңіз? (Қажет жағдайда бірнеше жауап белгілесеңіз болады.)<br>☐ Басқа интернет сайттағы сілтемесі арқылы<br>☐ Іздеу жүйесін қолдану арқылы<br>☐ Газет<br>☐ Телевизия<br>☐ Радио<br>☐ Достар/Жанұя мүшелері<br>☐ Басқасы, нұсқаңыз ________________________ | | | | | | | |
| 7 | Қазақстан «Электрондық үкімет» порталында болдыңыз ба?<br>☐ Иә<br>☐ Жоқ (17-ші сұраққа барыңыз) | | | | | | | |
| 8 | Сіз аптасына Қазақстан «Электрондық үкімет» порталында қандай жиілікпен боласыз? (Бір жауапты таңдаңыз.)<br>☐ 0<br>☐ 1 – 5 рет<br>☐ 6 – 9 рет<br>☐ 10 рет немесе одан да көп | | | | | | | |
| Келесі тұжырымдарға байланысты ойыңызды мынандай үлгіде беріңіз. Әрбір жауап нұскада жеті | | | | | | | | |
| **№** | **Сұрақ** | Толық келісемін | Орташа келісемін | Аздап келісемін | Бейтарап | Аздап келіспеймін | Орташа келіспеймін | Мүлдем келіспеймін |
| 9 | Қазақстан «Электрондық үкімет» порталының мазмұнын _________ деп табамын. | | | | | | | |
| | жаңартылған | | | | | | | |
| | толық | | | | | | | |
| | нақты/сенімді | | | | | | | |
| | баянсыз | | | | | | | |
| | пайдалы | | | | | | | |
| 10 | Мен ___________ деп табамын. | | | | | | | |
| | порталды түсінуге оңай | | | | | | | |
| | порталды ақпаратты іздеуде оңай | | | | | | | |
| | порталды түршіктіретін | | | | | | | |
| | порталды пайдаланған кезде, маған қосымша көмек қажет | | | | | | | |
| | порталдағы аса көп сілтеме жаңылдыратын | | | | | | | |
| | порталды колдануда оңай | | | | | | | |
| 11 | Мен келесі сектордағы ақпаратты пайдалы деп табамын | | | | | | | |
| | Білім | | | | | | | |
| | Денсаулық | | | | | | | |
| | Еңбек | | | | | | | |
| | Мәдениет | | | | | | | |
| | Спорт | | | | | | | |
| | Туризм | | | | | | | |
| 12 | Мен  төмендегі мемлекеттік және жеке меншік секторындағы қызметтер туралы ақпаратты пайдалы деп табамын. | | | | | | | |
| | Әуекомпаниялар | | | | | | | |
| | Автокөлік | | | | | | | |
| | Әуежайлар | | | | | | | |
| | Темір жол вокзалдары | | | | | | | |
| | Қонақ үйлер каталогы | | | | | | | |
| | Жүктасымалдау | | | | | | | |
| | Пошта | | | | | | | |
| | Интернет | | | | | | | |
| | Телекоммуникациялар | | | | | | | |
| | Жұмысқа орналастыру | | | | | | | |
| | Қалалардағы анықтамалық қызметтер | | | | | | | |
| | Заңдық қызметтер | | | | | | | |
| 13 | Мен келесі сектордағы маған кажет ақпаратты жеңіл деп табамын | | | | | | | |
| | Білім | | | | | | | |
| | Денсаулық | | | | | | | |
| | Еңбек | | | | | | | |
| | Мәдениет | | | | | | | |
| | Спорт | | | | | | | |
| | Туризм | | | | | | | |
| 14 | Мен келеси мемлекеттік және жеке меншік секторындағы төмендегі қызметтер туралы ақпаратты қолдануды оңай деп табамын | | | | | | | |
| | Әуекомпаниялар | | | | | | | |
| | Автокөлік | | | | | | | |
| | Әуежайлар | | | | | | | |
| | Темір жол вокзалдары | | | | | | | |
| | Қонақ үйлер каталогы | | | | | | | |
| | Жүктасымалдау | | | | | | | |
| | Пошта | | | | | | | |
| | Интернет | | | | | | | |
| | Телекоммуникациялар | | | | | | | |
| | Жұмысқа орналастыру | | | | | | | |
| | Қалалардағы анықтамалық қызметтер | | | | | | | |
| | Заңдық қызметтер | | | | | | | |
| 15 | Мен порталды қолдануды жоспарлап отырмын, себебі ол ________ үшін көмектеседі. | | | | | | | |
| | Шығын азайту | | | | | | | |
| | Кетіретін уақытты азайту | | | | | | | |
| | Бюрократияны азайту | | | | | | | |
| | Соңғы ақпаратты және жаналықтар алу | | | | | | | |
| | Менім сұрауыма жылдам жылдам жауап алу | | | | | | | |
| 16 | Мен басқа деректер көзін қолданғандықтан (ресми емес вэб ресурстары, ресми вэб ресурстары және СМИ) мен порталмен қолданбаймын | | | | | | | |
| **№** | **Сұрақ** | | | | | | | |
| 17 | Порталды қолданбауыңыздың себебтерін белгілеңіз. (Қажет жагдайда бірнеше жауап белгілесеңіз болады.)<br>☐ Мен пайдалы ақпарат таба алмаймын<br>☐ Порталда “Көмек” жоқ<br>☐ Портал өте ақырын жұмыс істейді<br>☐ Порталмен жұмыс жасау ыңғайсыз<br>☐ Мен порталдағы ақпаратқа сенбеймін<br>☐ Менде интернетке (жиі) шығатын мукіншілік жоқ.<br>☐ Мен порталды қолдануды қажет етпеймін<br>☐ Басқасы, нұсқаңыз ________________________ | | | | | | | |

| № | Демографикалық сұрақ | |
|---|---|---|
| 18 | Жыныс | ☐ Ер ☐ Әйел |
| 19 | Жас | ☐ 16 – 25 ☐ 26 – 35<br>☐ 36 – 55 ☐ 56 және жоғары |
| 20 | Білім деңгейі | ☐ Орта ☐ Бакалавр<br>☐ Магистратура ☐ Докторантура |
| 21 | Тұрғылықты жеріңіз | ☐ Алматы ☐ Астана<br>☐ Басқасы,нұсқаңыз______________ |
| 22 | Жұмыс түрі | ☐ Мемлекет қызметкер ☐ Бизнес/ Финанс<br>☐ Инженер ☐ Ақпарттық технологиялар<br>☐ Сауда ☐ Студент<br>☐ Жұмыссыз<br>☐ Басқасы, нұсқаңыз__________________ |
| 23 | Жалақы | ☐ 15000 тенгеден ($100) кем<br>☐ 15001-75000 тенге ($100-$500)<br>☐ 75001-150000 тенге ($501-$1000)<br>☐ 150001-300000 тенге ($1001-$2000)<br>☐ 300000 тенгеден ($2001) жоғары |

**Questionnaire in Russian**

Мы группа студентов университета НТУ, в данное время работаем над дипломом по теме восприятия пользователями единного портала электронного правительства Республики Казахстан. Информация, полученая нами, будет использоваться только в целях академического иследования. Все ответы будут строго конфеденциальны. Ваше участие добровольно, Вы можете отказаться в любое время.
Мы благодарны Вам за Ваше время и терпение, пожалуйста отвечайте на вопросы открыто и честно.
Спасибо, Балгаев Ильяс, Аманбек Ерлан, Батырханов Канат

{w070057, yerl0001, baty0001} @ntu.edu.sg

| № | Вопрос |
|---|---|
| 1 | Как часто Вы пользуетесь интернетом в неделю? (Выберите один вариант)<br>☐ Меньше чем 1 час<br>☐ 1 – 4 часов<br>☐ 5 – 9 часов<br>☐ 10 часов и больше |
| 2 | Откуда Вы обычно выходите в интернет? (Выберите один вариант)<br>☐ С работы<br>☐ Из дома<br>☐ Из интернет кафе<br>☐ Другое, пожалуйста укажите ________________________ |
| 3 | Какой у Вас тип интернет соединения? (Выберите один вариант)<br>☐ Аналоговая линия<br>☐ Широкополосный<br>☐ Спутниковый |
| 4 | Сколько Вы платите за интернет? (Выберите один вариант)<br>☐ 0 – 4500 тенге (US$30)<br>☐ 4501 тенге (US$30) – 7500 тенге (US$50)<br>☐ 7501 тенге (US$51) и выше |
| 5 | Слышали ли Вы о едином портале элетронного правительства Республики Казахстан?<br>☐ Да<br>☐ Нет (Пожалуйста перейдите к вопросу 18) |
| 6 | Как Вы узнали о едином портале элетронного правительства Республики Казахстан? (При необходимости, выберите несколько вариантов)<br>☐ Ссылка с интернет ресурса<br>☐ Использовали поисковую машину<br>☐ Газета<br>☐ ТВ<br>☐ Радио<br>☐ Друзья/Члены семьи<br>☐ Другое, пожалуйста укажите ________________________ |
| 7 | Посещали ли Вы единый портал элетронного правительства Республики Казахстан?<br>☐ Да<br>☐ Нет (Пожалуйста перейдите к вопросу 17) |
| 8 | Как часто Вы посещаете единый портал элетронного правительства Республики Казахстан в неделю? (Выберите один вариант)<br>☐ 0<br>☐ 1 – 5 раз<br>☐ 6 – 9 раз<br>☐ 10 раз и чаще |

Следующие вопросы будут заданы в формате, где будет учитываться Ваше мнение

| № | Вопрос | Категорически согласен | Умеренно согласен | Слегка согласен | Нейтрально | Слегка несогласен | Умеренно несогласен | Категорически есогласен |
|---|---|---|---|---|---|---|---|---|
| 9 | Содержимое единного портала элетронного правительства Республики Казахстан я нахожу: | | | | | | | |
| | свежим/обновленным | | | | | | | |
| | законченным/завершенным | | | | | | | |
| | аккуратным/достоверным | | | | | | | |
| | неповторяющимся | | | | | | | |
| | полезным | | | | | | | |
| 10 | Я нахожу портал легко понятным | | | | | | | |
| | Я нахожу поиск информации по порталу легким | | | | | | | |
| | Я нахожу работу с порталом раздражающей | | | | | | | |
| | Я считаю, что нуждаюсь в помощи при работе с порталом | | | | | | | |
| | Я считаю, что существует слишком много ссылок, которые меня смущают | | | | | | | |
| | Я нахожу портал полезным | | | | | | | |
| 11 | Я нахожу информацию в секторах, приведенных ниже, полезной | | | | | | | |
| | Образование | | | | | | | |
| | Здоровье | | | | | | | |
| | Труд | | | | | | | |
| | Культура | | | | | | | |
| | Спорт | | | | | | | |
| | Туризм | | | | | | | |
| 12 | Я нахожу информацию об услугах государственного и частного сектора, приведенных ниже, полезной | | | | | | | |
| | Авиакомпании | | | | | | | |
| | Автотранспорт | | | | | | | |
| | Аэропорты | | | | | | | |
| | Железнодорожные вокзалы | | | | | | | |
| | Каталог гостиниц | | | | | | | |
| | Грузоперевозки | | | | | | | |
| | Почта | | | | | | | |
| | Интернет | | | | | | | |
| | Телекоммуникации | | | | | | | |
| | Трудоустройство | | | | | | | |
| | Справочные службы городов | | | | | | | |
| | Юридические услуги | | | | | | | |
| 13 | Я могу легко найти требуемую информацию в данных секторах | | | | | | | |
| | Образование | | | | | | | |
| | Здоровье | | | | | | | |
| | Труд | | | | | | | |
| | Культура | | | | | | | |
| | Спорт | | | | | | | |
| | Туризм | | | | | | | |
| 14 | Я нахожу использование информации об услугах государственного и частного сектора, приведенных ниже, легкой | | | | | | | |
| | Авиакомпании | | | | | | | |
| | Автотранспорт | | | | | | | |
| | Аэропорты | | | | | | | |
| | Железнодорожные вокзалы | | | | | | | |
| | Каталог гостиниц | | | | | | | |
| | Грузоперевозки | | | | | | | |
| | Почта | | | | | | | |
| | Интернет | | | | | | | |
| | Телекоммуникации | | | | | | | |
| | Трудоустройство | | | | | | | |
| | Справочные службы городов | | | | | | | |
| | Юридические услуги | | | | | | | |
| 15 | Я планирую пользоваться порталом, потому что считаю, что это поможет мне | | | | | | | |
| | Снизить расходы | | | | | | | |
| | Снизить затрачиваемое время | | | | | | | |
| | Уменьшить бюрократию | | | | | | | |
| | Получить последнюю информацию и новости | | | | | | | |
| | Получить быстрые ответы на мои запросы | | | | | | | |
| 16 | Я не буду пользоваться порталом, т.к. уже пользуюсь другими источниками (неофициальные вэб ресурсы, официальные вэб ресурсы и каналы СМИ) | | | | | | | |

| № | Вопрос |
|---|---|
| 17 | Какие причины заставляют Вас не пользоваться единым порталом электронного правительства Республики Казахстан? (При необходимости, выберите несколько вариантов)<br>☐ Я не могу найти полезную информацию<br>☐ На портале нету "Помощи"<br>☐ Портал работает слишком медленно<br>☐ С порталом неудобно работать<br>☐ Я не доверяю информации на портале<br>☐ У меня нет доступа к интернету (часто)<br>☐ Я не чувствую необходимость в использовании портала<br>☐ Другое, пожалуйста укажите ________________________ |

**Демографический профайл**

| № | Вопрос | | |
|---|---|---|---|
| 18 | Пол | ☐ Муж | ☐ Жен |
| 19 | Возраст | ☐ 16 – 25<br>☐ 36 – 55 | ☐ 26 – 35<br>☐ 56 и выше |
| 20 | Образование | ☐ Среднее<br>☐ Магистр | ☐ Бакалавр<br>☐ Доктор |
| 21 | Место жительства | ☐ Алматы<br>☐ Другое, пожалуйста укажите_______ | ☐ Астана |
| 22 | Вид деятельности | ☐ Гос служба<br>☐ Инженер<br>☐ Торговля<br>☐ Безработный<br>☐ Другое, пожалуйста укажите_______ | ☐ Бизнес<br>☐ ИТ<br>☐ Студент |
| 23 | Заработная плата | ☐ меньше 15000 тенге ($100)<br>☐ 15001-75000 тенге ($100-$500)<br>☐ 75001-150000 тенге ($501-$1000)<br>☐ 150001-300000 тенге ($1001-$2000)<br>☐ Выше 300000 тенге ($2001) | |

**APPENDIX B**

| No. | Variable | Description of variable | Valid value |
|---|---|---|---|
| Q1 | MFRQINT | How frequently do you use the internet per week? | 1: Less than 1 hour<br>2: 1 – 4 hours<br>3: 5 – 9 hours<br>4: More than 10 hours |
| Q2 | MACCINT | Where do you usually access the internet? | 1:Work<br>2: Home<br>3: Internet cafe<br>0: other |
| Q3 | MCONTYP | What is your internet connection type? | 1: Dial-up<br>2: Broadband<br>3: Satellite |
| Q4 | MINTCOST | How much do you pay for internet connection? | 1: 0 – 4500 tenge (US$30)<br>2: 4501 tenge (US$30) – 7500 tenge (US$50)<br>3: 7501 tenge (US$51) and above |
| Q5 | MHEARD | Have you ever heard about the Kazakhstan e-Government portal? | 1: Yes<br>2: No |
| Q6 | ASOREGOV | How did you know about Kazakhstan e-Government portal? | 1: Link from other web-site<br>2: Internet search engine<br>3: Newspaper<br>4: TV<br>5: Radio<br>6: Friend/Family member<br>0: other |
| Q7 | AVISEGOV | Have you ever visited the Kazakhstan's e-Government portal? | 1: Yes<br>2: No |
| Q8 | AFRQEGOV | How many times do you visit the e-Government portal per week? | 1: Less than once<br>2: 1 – 5 times<br>3: 6 – 9 times<br>4: 10 times and above |
| Q9 | TPUCNTCUR | Current | 1: Strongly agree<br>2: Moderately agree<br>3: Slightly agree<br>4: Neutral<br>5: Slightly disagree<br>6: Moderately disagree |
| | | | 7: Strongly disagree |
| | TPUCNTCOM | Complete | Same as above |
| | TPUCNTACC | Accurate | Same as above |
| | TPUCNTNORED | Non redundant | Same as above |
| | TPUCNTU | Useful | Same as above |
| Q10 | TPEU | With reference to the Kazakhstan e-Government portal I find... | |
| | TPEUEUN | it easy to understand | Same as above |
| | TPEUESI | it easy to search for information | Same as above |
| | TPEUIF | it frustrating | Same as above |
| | TPEUHLP | I need help while using it | Same as above |
| | TPEUCONF | there are too many links that confuse me | Same as above |
| | TPEUEU | it easy to use | Same as above |
| Q11 | TPU | I find the information in following sectors useful: | |
| | TPUEDUC | Education | Same as above |
| | TPUHLTH | Health | Same as above |
| | TPUEMPL | Employment | Same as above |
| | TPUCULT | Culture | Same as above |
| | TPUSPT | Sport | Same as above |
| | TPUTORS | Tourism | Same as above |
| Q12 | TPU | I find the Information on following government and private services useful: | |
| | TPUAIRL | Airlines | Same as above |
| | TPUMTRAN | Motor transport | Same as above |
| | TPUAIRP | Airports | Same as above |
| | TPURAIL | Railroad stations | Same as above |
| | TPUHOT | Hotels | Same as above |
| | TPUTRAN | Transportation | Same as above |
| | TPUPOS | Post | Same as above |
| | TPUINT | Internet | Same as above |
| | TPUTEL | Telecommunications | Same as above |
| | TPUEMP | Employment | Same as above |
| | TPUMRS | Municipal reference services | Same as above |
| | TPULS | Legal services | Same as above |
| Q13 | TPEU | I can find relevant and appopriate information that I need easily in the following sectors: | |
| | TPEUEDUC | Education | Same as above |
| | TPEUHLTH | Health | Same as above |
| | TPEUEMPL | Employment | Same as above |
| | TPEUCULT | Culture | Same as above |
| | TPEUSPT | Sport | Same as above |
| | TPEUTORS | Tourism | Same as above |
| Q14 | TPEU | I find using following information on government and private services easy to use: | |
| | TPEUAIRL | Airlines | Same as above |
| | TPEUMTRAN | Motor transport | Same as above |
| | TPEUAIRP | Airports | Same as above |
| | TPEURAIL | Railroad stations | Same as above |
| | TPEUHOT | Hotels | Same as above |
| | TPEUTRAN | Transportation | Same as above |
| | TPEUPOS | Post | Same as above |
| | TPEUINT | Internet | Same as above |
| | TPEUTEL | Telecommunications | Same as above |
| | TPEUEMP | Employment | Same as above |
| | TPEUMRS | Municipal reference services | Same as above |
| | TPEULS | Legal services | Same as above |
| Q15 | TBI | I would like to use e-Government web portal because it helps me: | |
| | TBIREDXP | To reduce my expenses | Same as above |
| | TBIREDTM | To reduce my time spent | Same as above |
| | TBIMINBUR | To minimize bureaucracy | Same as above |
| | TBILTSINF | To get latest information and news | Same as above |
| | TBUQRES | To get quick responses for inquiries | Same as above |
| Q16 | TBIALT | As I am currently using other sources for the information I need (e.g. informal web sites, official resources and mass media channels) | Same as above |
| Q17 | ABARRIER | What are the reasons for not using/visiting Kazakhstan e-Government web portal? | Same as above |
| Q18 | DGEN | Gender | 1: Male<br>2: Female |
| Q19 | DAGE | Age | 1: 16 – 25<br>2: 26 – 35<br>3: 36 – 55<br>4: 56 and above |
| Q20 | DACCLVL | Academic level | 1: High school<br>2: University Bachelor<br>3: University Master<br>4: University PhD |
| Q21 | DPLA | Place of staying | 1: Almaty<br>2: Astana<br>0: other |
| Q22 | DOCCP | Occupation | 1: Government<br>2: Finance<br>3: Engineering<br>4: IT<br>5: Sale<br>6: Student<br>7: Unemployed<br>0: other |
| Q23 | DINCM | Salary range | 1: under 15000 tenge ($100)<br>2: 15001-75000 tenge ($100-$500)<br>3: 75001-150000 tenge($501-$1000)<br>4: 150001-300000tenge ($1001-$2000)<br>5: Above 300000 tenge ($2000) |

**Table B1. Data coding.**

## APPENDIX C

| Hypothesis | Variable | Type of variable | Approach |
|---|---|---|---|
| H1 | PU | Numeric | Linear regression |
| | PEOU | Numeric | |
| H2 | PU | Numeric | Linear regression |
| | BI | Numeric | |
| H3 | PU | Numeric | Linear regression |
| | BI | Numeric | |
| H4 | BI | Numeric | Independent Sample T-test |
| | AU | Categorical | |
| H5 | BI | Numeric | Linear regression |
| | External factors: Occupation, Academic level | Categorical | |

**Table C1. Summary of approaches for hypotheses testing.**

| Variables description | Component |
|---|---|
| | 1 |
| With reference to the Kazakhstan e-Government portal I find the content: current | 0.393 |
| With reference to the Kazakhstan e-Government portal I find the content: complete | 0.371 |
| With reference to the Kazakhstan e-Government portal I find the content: accurate | 0.298 |
| With reference to the Kazakhstan e-Government portal I find the content: non redundant | 0.502 |
| With reference to the Kazakhstan e-Government portal I find the content: useful | 0.565 |
| I find the information in following sectors useful: Education | 0.814 |
| I find the information in following sectors useful: Health | 0.822 |
| I find the information in following sectors useful: Employment | 0.779 |
| I find the information in following sectors useful: Culture | 0.828 |
| I find the information in following sectors useful: Sport | 0.702 |
| I find the information in following sectors useful: Tourism | 0.694 |
| I find the Information on following government and private services useful: Airlines | 0.913 |
| I find the Information on following government and private services useful: Motor transport | 0.88 |
| I find the Information on following government and private services useful: Airports | 0.921 |
| I find the Information on following government and private services useful: Railroad stations | 0.906 |
| I find the Information on following government and private services useful: Hotels | 0.897 |
| I find the Information on following government and private services useful: Transportation | 0.835 |
| I find the Information on following government and private services useful: Post | 0.9 |
| I find the Information on following government and private services useful: Internet | 0.919 |
| I find the Information on following government and private services useful: Telecommunications | 0.934 |
| I find the Information on following government and private services useful: Employment | 0.862 |
| I find the Information on following government and private services useful: Municipal reference services | 0.912 |
| I find the Information on following government and private services useful: Legal services | 0.894 |

**Table C2. Component matrix of PU constructs.**

| Variables description | Component |
|---|---|
| | 1 |
| With reference to the Kazakhstan e-Government portal I find...it easy to understand | 0.587 |
| With reference to the Kazakhstan e-Government portal I find...it easy to search for information | 0.644 |
| With reference to the Kazakhstan e-Government portal I find...it frustrating | 0.084 |
| With reference to the Kazakhstan e-Government portal I find...I need help while using it | 0 |
| With reference to the Kazakhstan e-Government portal I find...there are too many links that confuse me | -0.06 |
| With reference to the Kazakhstan e-Government portal I find...it easy to use | 0.572 |
| I can find relevant and appopriate information that I need easily in the following sectors: Education | 0.791 |
| I can find relevant and appopriate information that I need easily in the following sectors: Health | 0.75 |
| I can find relevant and appopriate information that I need easily in the following sectors: Employment | 0.737 |
| I can find relevant and appopriate information that I need easily in the following sectors: Culture | 0.789 |
| I can find relevant and appopriate information that I need easily in the following sectors: Sport | 0.766 |
| I can find relevant and appopriate information that I need easily in the following sectors: Tourism | 0.785 |
| I find using following information on government and private services easy to use: Airlines | 0.909 |
| I find using following information on government and private services easy to use: Motor transport | 0.917 |
| I find using following information on government and private services easy to use: Airports | 0.911 |
| I find using following information on government and private services easy to use: Railroad stations | 0.918 |
| I find using following information on government and private services easy to use: Hotels | 0.935 |
| I find using following information on government and private services easy to use: Transportation | 0.9 |
| I find using following information on government and private services easy to use: Post | 0.903 |
| I find using following information on government and private services easy to use: Internet | 0.922 |
| I find using following information on government and private services easy to use: Telecommunications | 0.929 |
| I find using following information on government and private services easy to use: Employment | 0.892 |
| I find using following information on government and private services easy to use: Municipal reference services | 0.896 |
| I find using following information on government and private services easy to use: Legal services | 0.925 |

**Table C3. Component matrix of PEOU constructs.**

| | Component |
|---|---|
| | 1 |
| I would like to use e-Government web portal because it helps me: To reduce my expenses | 0.862 |
| I would like to use e-Government web portal because it helps me: To reduce my time spent | 0.847 |
| I would like to use e-Government web portal because it helps me: To minimize bureaucracy | 0.904 |
| I would like to use e-Government web portal because it helps me: To get latest information and news | 0.871 |
| I would like to use e-Government web portal because it helps me: To get quick responses for inquiries | 0.887 |
| Although I am currently using other sources for the information I need (e.g. informal web sites, official resources and mass media channels) I would use the portal | 0.166 |

**Table C4. Component matrix of BI constructs.**

## APPENDIX D

| No. | Model | R-square | B | Sig. | Result |
|---|---|---|---|---|---|
| H1 | PU = PEOU + error | 0.542 | | | |
| | PEOU | | 0.685 | p < 0.01 | Accepted |
| H2 | BI = PU + error | 0.199 | | | |
| | PU | | 0.475 | p < 0.01 | Accepted |
| H3 | BI = PEOU + error | 0.048 | | | |
| | PEOU | | 0.228 | p < 0.165 | Rejected |
| H5 | BI = PU + PEOU + Ocupation + Academic level + error | | | | |
| | | 0.477 | | | |
| | PU | | 0.477 | p < 0.01 | |
| | PEOU | | 0 | p < 0.36 | |
| | Ocupation | | -0.155 | p < 0.05 | |
| | Academic level | | 0 | p < 0.862 | Rejected |

**Table D1. Results of hypotheses testing (H1, H2, H3, H5).**

| No. | Model | Levene's Test for Equality of Variances | | Sig. | Result |
|---|---|---|---|---|---|
| | | F | Sig. | | |
| H4 | AU = BI + error | 1.32 | 0.25 | p < 0.139 | Rejected |

**Table D2. Result of Hypothesis 4 (Independent Sample T-test).**

**APPENDIX E**

| Sectors | PU | |
|---|---|---|
| | **Mean** | **Std. Deviation** |
| Education | 2.6 | 1.321 |
| Employment | 2.68 | 1.371 |
| Health | 2.73 | 1.394 |
| Culture | 2.98 | 1.374 |
| Sport | 3.15 | 1.377 |
| Tourism | 3.21 | 1.392 |

**Table E1. Ranking of information sectors by mean score of PU.**

| Services | PU | |
|---|---|---|
| | **Mean** | **Std. Deviation** |
| Airports | 2.98 | 1.537 |
| Municipal reference services | 2.98 | 1.705 |
| Railroad stations | 3.03 | 1.553 |
| Airlines | 3.05 | 1.49 |
| Motor transport | 3.08 | 1.535 |
| Telecommunications | 3.09 | 1.654 |
| Post | 3.1 | 1.528 |
| Internet | 3.1 | 1.626 |
| Legal services | 3.12 | 1.68 |
| Hotels | 3.21 | 1.472 |
| Employment | 3.28 | 1.641 |
| Transportation | 3.33 | 1.538 |

**Table E2. Ranking for information about services by mean score of PU.**

| Sectors | PEOU | |
|---|---|---|
| | Mean | Std. Deviation |
| Education | 2.95 | 1.301 |
| Health | 3.16 | 1.554 |
| Employment | 3.24 | 1.401 |
| Culture | 3.36 | 1.257 |
| Sport | 3.39 | 1.317 |
| Tourism | 3.39 | 1.386 |

**Table E3. Ranking of information sectors by mean score of PEOU.**

| Services | PEOU | |
|---|---|---|
| | Mean | Std. Deviation |
| Airlines | 3.4 | 1.334 |
| Motor transport | 3.41 | 1.332 |
| Airports | 3.37 | 1.358 |
| Railroad stations | 3.37 | 1.397 |
| Hotels | 3.52 | 1.348 |
| Transportation | 3.59 | 1.247 |
| Post | 3.45 | 1.361 |
| Internet | 3.43 | 1.386 |
| Telecommunications | 3.44 | 1.376 |
| Employment | 3.6 | 1.314 |
| Municipal reference services | 3.49 | 1.526 |
| Legal services | 3.46 | 1.414 |

**Table E4. Ranking for information about services by mean score of PEOU.**